\documentstyle[prc,preprint,epsfig,aps]{revtex}
\newcommand{\ga}{\alpha}
\newcommand{\gb}{\beta}
\newcommand{\gc}{\gamma}
\newcommand{\gd}{\delta}

\newcommand{\gve}{\varepsilon}
\newcommand{\gl}{\lambda}

\newcommand{\gS}{\Sigma}

\newcommand{\mbf}[1]{\mbox{\boldmath$#1$}}

\tightenlines
\begin{document}


\title{Medium corrections in the formation of light charged
particles in   heavy ion reactions}

\author{ C. Kuhrts$^1$, M. Beyer$^1$\footnote{corresponding author: FB Physik,
 Universit\"at Rostock, Universit\"atsplatz 3,
18051 Rostock, Germany, Phone:
[+49] 381/498-1674, Fax: -1673,
 email:michael.beyer@physik.uni-rostock.de},
P. Danielewicz$^2$ and G. R\"opke$^1$
}
\address{
$^1$FB Physik,
 Universit\"at Rostock, Universit\"atsplatz 3,
18051 Rostock, Germany\\
$^2$ NSCL, Michigan State University, East Lansing, MI 48824, U.S.A.}

\vspace{10mm}

\maketitle

\begin{abstract}
  Within a microscopic statistical description of heavy ion
  collisions, we investigate the effect of the medium on the formation
  of light clusters.  The~dominant medium effects are self-energy
  corrections and Pauli blocking that produce the Mott effect for
  composite particles and enhanced reaction rates in the collision
  integrals.  Microscopic description of composites in the medium
  follows the Dyson equation approach combined with the cluster
  mean-field expansion.  The~resulting effective few-body problem is
  solved within a properly modified Alt-Grassberger-Sandhas formalism.
  The~results are incorporated in a Boltzmann-Uehling-Uhlenbeck
  simulation for heavy ion collisions.  The number and spectra of
  light charged particles emerging from a heavy ion collision
  changes in a~significant manner in effect of the medium modification
  of production and absorption processes.
\end{abstract}

\vspace{5mm}

\section{Introduction}
The description of the dynamics of an interacting many-body system is
particularly difficult when the quasiparticle approach reaches its
limits.  That may be the case when the residual interaction is strong
enough to build up correlations.  An example of such a system with
correlations is nuclear matter in heavy-ion collisions.
In~collisions, the nuclear matter is first compressed and excited and
then decompressed.  At another extreme of macroscopic scale, the
nuclear matter is formed during a supernova collapse and becomes the
material of which a neutron star is made.

Here, we address the simplest case of the formation of correlations in
a~nonequilibrium situation: light-cluster production in low density
nuclear matter.  The~production can be described microscopically
following the set of generalized semiclassical Boltzmann equations.
In~the past, the~coupled set was solved numerically for nucleon
$f_N$, deuteron $f_d$, triton $f_t$, and helium-3 $f_h$ Wigner
functions within the Boltzmann-Uehling-Uhlenbeck (BUU)
approach~\cite{dan91,Dan92}.  The formation and disintegration of
clusters generally takes place in reactions involving different
particles.
The~rates for respective processes are utilized in the
Boltzmann collision integrals.

Rates for collisions, whether involving clusters or not, are
central ingredients of all modern
microscopic approaches to heavy reactions such as the
Boltzmann-Uehling-Uhlenbeck (BUU)
approach~\cite{dan91,Dan92,sto86,wolter} or the quantum molecular
dynamics (QMD)~\cite{QMD,QMD2}.  For the status of those approaches
see Ref.~\cite{fel99}.

The formation and disintegration of deuterons are
the simplest types of processes involving clusters discussed
above.  Notably,
with the exception of electromagnetic production and breakup
($np\rightleftharpoons \gamma d$), these processes involve the minimum
of three elementary particles ($NNN\rightleftharpoons Nd$) and,
correspondingly, require the solution of a~three-body problem in the
medium in their description.  Complication brought in by the medium is
evident even in the kinematics, as the medium brings in a~preferred
frame different from the center of mass of the three bodies.

In general, the~collision rates are affected by the surrounding
medium.  The~common procedure is to ignore the medium dependence and
to utilize free-space {\em experimental cross sections}.  That
procedure in many cases was very successful.  To calculate the rates,
including the {\em self energy shift} and the proper {\em Pauli
blocking}, and to study the influence of the medium on different
observables, a~generalized Alt-Grassberger-Sandhas (AGS)
equation~\cite{AGS} has been
derived~\cite{bey96,habil,bey97,beyFB,schadow,kuhrts,alpha}.
The~effective few-body problem in matter arises within the Green
function method~\cite{fet71} when following the cluster mean-field
expansion~\cite{RMSS} or the Dyson equation approach~\cite{duk98}.
Besides the medium-dependent cross-sections, the~Mott
effect~\cite{RMOTT1,RMOTT2} plays an~important role in nuclear matter.
Both effects are studied in this paper in the context of a~realistic
simulation of a~heavy ion reaction.  As~an example we choose
the
reaction studied in a recent experiment by the INDRA Collaboration,
namely $^{129}$Xe+$^{119}$Sn at 50 MeV/A~\cite{INDRA}.

The BUU approach will be explained in the following section.
The~needed three-body reaction input from the AGS-type equations will
be discussed in Sec.~\ref{sec:green}.  Details of the derivation of
these equations within the Green function approach may be e.g. found
in~\cite{fet71}.  In Sec.~\ref{sec:result} we shall present numerical
results and in the last section we shall summarize the paper and give
conclusions.

\section{Nuclear Reactions}

Within the microscopic statistical approach, the~reacting system is
described in terms of the Wigner distributions $f_X$ for light
particles, thus for nucleon $f_N$, deuteron $f_d$, triton $f_t$,
and helion $f_h$.  Within the cluster mean-field
approximation, these distributions follow the set of coupled
semiclassical Boltzmann equations~\cite{dan91},
\begin{equation}
\partial_t f_X+\{{\cal U}_X,f_X\}=
{\cal K}^{\rm gain}_X \lbrace f_N,f_d,f_t,\dots \rbrace \,
(1\pm f_X)
-{\cal K}^{\rm loss}_X \lbrace f_N,f_d,f_t,\dots \rbrace \,
f_X, \quad X=N,d,t,\dots \label{eqn:Boltz}
\end{equation}
In the above, ${\cal U}_X$ is a mean-field potential and
$\{\cdot,\cdot\}$ denotes the Poisson bracket.  The~upper sign on the
r.h.s.\ is for the Bose and the lower for the Fermi type fragments.
Conversion between different species is accounted for in the collision
integrals ${\cal K} \lbrace f_N,f_d,f_t,\dots \rbrace $.  For example,
the~deuteron gain rate ${\cal K}^{\rm gain}_d(P,t)$ is
\begin{eqnarray}
{\cal K}^{\rm gain}_d(P,t)&=&
\int d^3k\int d^3k_1d^3k_2\; |\langle kP|U|k_1k_2\rangle|^2_
{dd\rightarrow dd}\;
f_d(k_1,t)f_d(k_2,t) \bar f_d(k,t)\nonumber\\
&&+\int d^3k\int d^3k_1d^3k_2\;
|\langle kP|U|k_1k_2\rangle|^2_{Nd\rightarrow Nd}\;
f_N(k_1,t)f_d(k_2,t) \bar f_N(k,t)\nonumber\\
&&+\int d^3k\int d^3k_1d^3k_2d^3k_3\;
|\langle kP|U_0|k_1k_2k_3\rangle|^2_{pnN\rightarrow dN}\nonumber\\&&
\qquad\times
f_N(k_1,t)f_N(k_2,t)f_N(k_3,t)\bar f_N(k,t)
+\dots\label{eqn:react}
\end{eqnarray}
Here we use the abbreviations $\bar f_N=(1-f_N)$ and $\bar
f_d=(1+f_d)$ and did not write overall energy-momentum
conserving $\delta$-functions.  The~ellipsis on the r.h.s.\ of
(\ref{eqn:react}) denote further possible contributions from four and
more body collisions (e.g.\ $tp\rightleftharpoons dd$,
$hp\rightleftharpoons dd$) or electromagnetic ($np\rightleftharpoons
\gamma d$).  The~loss rate ${\cal K}^{\rm loss}_d(P,t)$ is given by
\begin{eqnarray}
{\cal K}^{\rm loss}_d(P,t)&=&
\int d^3k\int d^3k_1d^3k_2d^3k_3\;
|\langle k_1k_2k_3|U_0|kP\rangle|^2_{dN\rightarrow pnN}\nonumber\\&&
\qquad\times
\bar f_N(k_1,t)\bar f_N(k_2,t)\bar f_N(k_3,t)f_N(k,t)
+\dots\label{eqn:react2}
\end{eqnarray}

The quantity $U_0$ appearing in Eqs.~(\ref{eqn:react}) and
(\ref{eqn:react2}) is the in-medium break-up transition operator for
the $Nd\rightarrow NNN$ reaction and it is calculated using the AGS
equation, as discussed in the next section.  For three nucleons
in
isolation, $U_0$~determines free-space break-up cross section
$\sigma_{\rm bu}^0$ via the equation~\cite{AGS,glo88}
\begin{eqnarray}
\sigma_{\rm bu}^0 (E)&=& \frac{1}{|v_d-v_N|} \frac{1}{3!}
\int d^3k_1d^3k_2d^3k_3\;
 |\langle kP|U_0|k_1k_2k_3\rangle|^2\nonumber\\
&&\qquad\qquad\times2\pi\delta(E'-E)\;(2\pi)^3\delta^{(3)}(k_1+k_2+k_3),
\label{eqn:sig0}
\end{eqnarray}
where $|v_d-v_N|$ denotes the relative velocity of the incoming
nucleon and deuteron. This equation along with detailed balance is
used to replace the squared transition matrix elements by the break-up
cross section~\cite{dan91}.  The~cross sections can be extracted from
data with theory aiding in interpolating and extrapolating.  This type
of procedure is widely used and has a~good chance of being successful
in the low density regime. However this procedure may not be
sufficient for heavy ion collisions particularly in the lower
energy
regime. The~elementary cross section that enters into the Boltzmann
equation depends on the medium, e.g.\ through blocking of intermediate
states in scattering or through self-energy for intermediate
states.
This has been demonstrated for three-nucleon processes in
Refs.~\cite{bey96,bey97}.

\section{Finite temperature few-body equations}
\label{sec:green}

The effective few-body problem in matter emerges within the
Green function
method~\cite{fet71} when following the cluster expansion~\cite{RMSS}
or related Dyson-equation approach~\cite{duk98}.  In the
cluster-mean field expansion, a~self-consistent system of equations
can be derived describing an $A$ particle cluster moving in a mean
field produced by the equilibrium mixture of clusters with different
particle number.  The~self-consistent determination of the composition
of the medium is a very challenging task that is not solved till now.
We adhere to an approximation where the correlations in the
medium
outside the considered cluster are neglected so that the
embedding
nuclear matter is described by the equilibrium distribution of nucleon
quasiparticles.

The formalism to describe three-body correlations in nuclear matter
has been discussed in detail
elsewhere~\cite{bey96,habil,bey97,beyFB,schadow,kuhrts,alpha}. Here,
we merely repeat some of the basic results for convenience.

Let the Hamiltonian of the system be given by
\begin{equation}
        H =\sum_{1} \frac{k_1^2}{2 m} a_1^\dagger a_{1}+\;\frac{1}{2}\;
        \sum_{12 1'2'}V_2(12,1'2')\;a^\dagger_1  a^\dagger_2  a_{2'}a_{1'}
\, ,
\label{eqn:H}
\end{equation}
where $a_1$ denotes the Heisenberg annihilation operator of the
particle with quantum numbers indexed by~1, including spin $s_1$ and
momentum~$k_1$.  The free resolvent $G_0$ for an $A$ particle cluster
is given in the Matsubara-Fourier representation by
\begin{equation}
        G_0(z) = (z-H_0)^{-1}\;N \equiv R_0(z)\;N \, .
\label{eqn:G0}
\end{equation}
In the above, $G_0$, $H_0$, and $N$ are all operators in the space of
$A$ particles.  For simplicity, an~index $A$ indicating that fact has
been dropped, but may get instituted when needed.  In~(\ref{eqn:G0}),
$z$~denotes the Matsubara frequencies $z_\gl=\pi\gl/(-i\gb)+\mu$ with
$\gl=0,\pm 2,\pm 4,\dots$ for bosons and $\gl=\pm 1,\pm 3,\dots$ for
fermions; $\mu$ is the chemical potential and $\gb=1/k_BT$ the inverse
temperature.  The effective in-medium Hamiltonian $H_0$ for the
quasi-particles in the mean field is given by
\begin{equation}
        H_0 = \sum_{i=1}^n
        \frac{k_i^2}{2m} +\Sigma_i\equiv \sum_{i=1}^n \varepsilon_i
\end{equation}
where the energy shift $\Sigma_1$ is
\begin{eqnarray}
        \Sigma_1&=&\sum_{2}V_2(12,\widetilde{12})f_2 \, ,
\label{eqn:selfHF}
\end{eqnarray}
with the
Fermi function $f_1$
\begin{eqnarray}
        f_1&\equiv& f(\varepsilon_1) = \frac{1}{\mbox{e}^{(\varepsilon_1 -
        \mu)/k_BT}+1} \, .
\label{eqn:fermi}
\end{eqnarray}
The tilde in $\widetilde{12}$ indicates antisymmetrization.
The factor
$N$ in Eq.~(\ref{eqn:G0}) is associated with particle statistics and
it is given by
\begin{equation}
        N=\bar f_1\bar f_2 \dots \bar f_A\pm f_1f_2\dots f_A
\end{equation}
where $\bar f=1-f$. The upper sign in $N$ is for an odd number of
constituent fermions and the lower sign is for an even number of
fermions or bosons.  Note that the commutation holds:
$NR_0=R_0N$.

The full resolvent $G$ in the Matsubara-Fourier representation may be
written as
\begin{equation}
        G(z)=(z-H_0-V)^{-1}N\equiv R(z)N \, ,
\label{eqn:G}
\end{equation}
where the potential $V$ is a sum over two-body interactions
within all possible pairs $\ga$ in the cluster, i.e.\
\begin{equation}
        V=\sum_\ga V_\ga= \sum_\ga N_2^{\ga}V_2^{\ga} \,  ,
\label{eqn:V2}
\end{equation}
where $V_2^{\ga}$ denotes the two-body potential in Eq.~(\ref{eqn:H}).
Note that, in consequence of Eq.~(\ref{eqn:V2}),
the~effective
potential is nonhermitian, $V^\dagger\neq V$, and also $R(z)N\neq
NR(z)$. This leads to the necessity of introducing the right and left
eigenvectors as e.g.\ done in the context of the four-body problem
in~\cite{alpha}.  For a pair $\ga=(12)$ of an $A$ particle cluster,
the~effective potential of Eq.~(\ref{eqn:V2}) is
\begin{equation}
\label{pauli}
        \langle 12|N_2^{(12)}V_2^{(12)}|1'2'\rangle = (\bar f_1\bar f_2 -
        f_1f_2)V_2(12,1'2').
\end{equation}
A useful operator is the channel resolvent $G_\ga(z)$ within an $A$
particle cluster, where only effective pair interaction in channel
$\ga$ is considered.  This operator may be represented as
\begin{eqnarray}
\nonumber
        G_\ga(z)&=&(z-H_0-V_\ga)^{-1}N\\
        &=&(z-H_0-N_2^\ga V_2^\ga)^{-1}N\equiv R_\ga(z)N.
\label{eqn:Ga}
\end{eqnarray}
Using inverses to the operators~$R$ defined in~(\ref{eqn:G0}),
(\ref{eqn:G}), and~(\ref{eqn:Ga}), it is possible to derive formally
the resolvent equations.  For the sake of similarity with the isolated
case, the~$A$ particle channel $t$ matrix $T_\ga$ is defined by
\begin{equation}\label{eqn:defRa}
        R_\ga(z)=R_0(z)+R_0(z)T_\ga(z) R_0(z)\,.
\end{equation}
Upon introducing $T_\ga(z)=N_2^\ga T_2^\ga(z)$,
Eq.~(\ref{eqn:defRa})
yields to the well-known Feynman-Galitzki equation~\cite{fet71}
\begin{equation}
        T_2^\ga(z)=V_2^\ga + V_2^\ga R_0(z) N_2^\ga  T_2^\ga (z).
\label{eqn:T2}
\end{equation}
We remark that equations of a similar form were derived by
different authors in the past~\cite{SchuckR,SchuckR1}.

Note that the above equations are also valid for the two-particle
subsystem embedded in a larger cluster (three, four, or more
particles).  As~for the isolated equations, the~effects of the
remaining particles appear only in the Matsubara frequencies $z$
(energies) of the considered particles within the cluster.  No
additional statistical factors related to the larger cluster arise.
One should note that the statistical effects generally arise in the
resolvent $G_0$ and not in the potential $V_2$. However, it is
possible to rewrite the respective equations introducing an effective
potential, as shown in Eq.~(\ref{eqn:T2}), and instead using unchanged
resolvents.  Following the more intuitive picture of the blocking in
the propagation of the particles, described by the resolvents, we
directly find in Eq.~(\ref{eqn:T2}) the~proper expression for the $t$
matrix which enters the Boltzmann collision integrals (see
e.g.~\cite{roepke}).

Derivation of the three-body in-medium equation is relatively
straightforward and was given in
Refs.~\cite{bey97,beyFB,schadow,kuhrts}.  The~AGS operator
$U_{\gb\ga}(z)$~\cite{AGS} for the three-particle system is
defined within the equation
\begin{equation}\label{eqn:defUtr}
        R(z)=\gd_{\gb\ga}R_\ga(z) + R_\gb(z) U_{\gb\ga}(z)
R_\ga(z) \, ,
\end{equation}
and it satisfies the following AGS-type equation
\begin{equation}
        U_{\gb\ga}(z)=\bar\gd_{\gb\ga}R_0(z)^{-1}+\sum_\gc\bar\gd_{\gb\gc}
        N_2^\gc T_2^\gc(z) R_0(z) U_{\gc\ga}(z) \, .
\label{eqn:AGS2}
\end{equation}
This equation includes consistently the medium effects, statistical
and self-energy shifts in the same way as the Feynman-Galitzki
equation for the two-particle correlations.  We use above the notation
$\bar\gd_{\ga\gb}= 1-\gd_{\ga\gb}$.  The~AGS-type equation yields the
three-body transition operator~$U$ for a~three-particle cluster as
well the transition operator for a~three-particle cluster embedded in
a~still larger cluster.  In~the latter case, the~effect of the other
particles in the cluster appears only in the Matsubara frequency
(energy)~$z$.  The~transition operator is defined in
Eq.~(\ref{eqn:defUtr}) in such a~manner that no additional factors $N$
accompany $U$ when $U$ is further employed. This in turn guarantees
that the cluster equations are valid also if they are part of a~larger
cluster.  The~two-body subsystem $t$ matrix entering in
Eq.~(\ref{eqn:AGS2}) is the same as the one given in
Eq.~(\ref{eqn:T2}).  In that way it is possible to use all results of
the few-body 'algebra' recurrently, in particular those based on the
cluster decomposition property.

The in-medium bound state equation for an $A$ particle cluster follows
from the homogeneous Lippmann-Schwinger type equation
\begin{equation}
        |\psi_B\rangle =R_0(E_B)V|\psi_B\rangle=
        R_0(E_B) \sum_{\gc} N_2^\gc V_2^\gc|\psi_B\rangle\,
\label{eqn:bound}
\end{equation}
where the sum is over all unique pairs in the cluster.
As shown in Ref.~\cite{schadow},
it is convenient to introduce the form factors
for the three-body bound state:
\begin{equation}
        |F_\gb\rangle=\sum_{\gc=1}^3\bar\gd_{\gb\gc} N_2^\gc  V_2^\gc
|\psi_{B_3}\rangle \, .
\end{equation}
With these, one derives
the homogeneous in-medium AGS-type equation
\begin{equation}
        |F_\ga\rangle=\sum_{\gb=1}^3 \bar\gd_{\ga\gb} N_2^\gb
        T_2^\gb R_0(B_3)|F_\gb\rangle.
\label{eqn:F3}
\end{equation}

Since the Green functions are evaluated in an~independent particle
basis, the~one-, \mbox{two-,} and three-particle Green functions are
decoupled in hierarchy.  To solve the in-medium problem up to
three-particle clusters, the one-, two-, and three-particle problems
are solved consistently.  The procedure generates
the single particle
self-energy shift, given by Eq.~(\ref{eqn:selfHF}), the
two-body input
including statistical factors in Eq.~(\ref{eqn:T2}), and, finally,
the three-body scattering state.

When solving the equations, technical reasons force us to adopt
some reasonable approximations valid at smaller effective densities.
Thus, the~nucleon self-energy is calculated via
Eq.~(\ref{eqn:selfHF}), but in the three-body equations we employ the
effective-mass approximation, i.e.\ we use
\begin{equation}
\gve_1=\frac{k_1^2}{2m}+\gS^{HF}(k_1)
\simeq \frac{k_1^2}{2m^*}+\gS^{HF}(P_{\rm c.m.}/3).
\end{equation}
For the sake of simplicity and the speed of the calculation, we
moreover utilize angle-averaged Fermi-function factors $\bar
N_2$, \begin{equation}
\bar N_2(p,q,P_{\rm c.m.}) =
\frac{1}{(4\pi)^2}\;\int d\cos\theta_q \; d\cos\theta_p \; d\phi_q
d\phi_p\; N_2(\mbf{p},\mbf{q},\mbf{P}_{\rm c.m.}) \, ,
\end{equation}
where $\mbf{p}$ and $\mbf{q}$ are the standard Jacobi
coordinates~\cite{glo88} and the angles are relative to
$\mbf{P}_{\rm c.m.}$.

Rank-one Yamaguchi potentials~\cite{yama} were employed in the
calculations of the deuteron break-up cross-section in the coupled
$^3$S$_1$-$^3$D$_1$ and the $^1$S$_0$ channels which should
suffice at low energies.  Through the dependence of the Fermi
functions on the center of mass momentum, the~integral equation for
the three-body system exhibits a dependence on the net momentum
in the medium frame, in sharp contrast to the free-space case.

\section{Results}\label{sec:result}

The coupled set of Boltzmann transport equations~\cite{dan91,Dan92},
for the Wigner distribution functions of nucleons~$f_N$,
deuterons~$f_d$, tritons~$f_t$, and helions~$f_h$, is solved for the
collisions of heavy ions.  For the clusters, treated in the
quasiparticle approximation, it is essential to consider the Mott
effect~\cite{dan91,RMOTT1,RMOTT2}. Without considering this effect the
description of data fails badly.

\begin{figure}[t]
  \begin{center}
\psfig{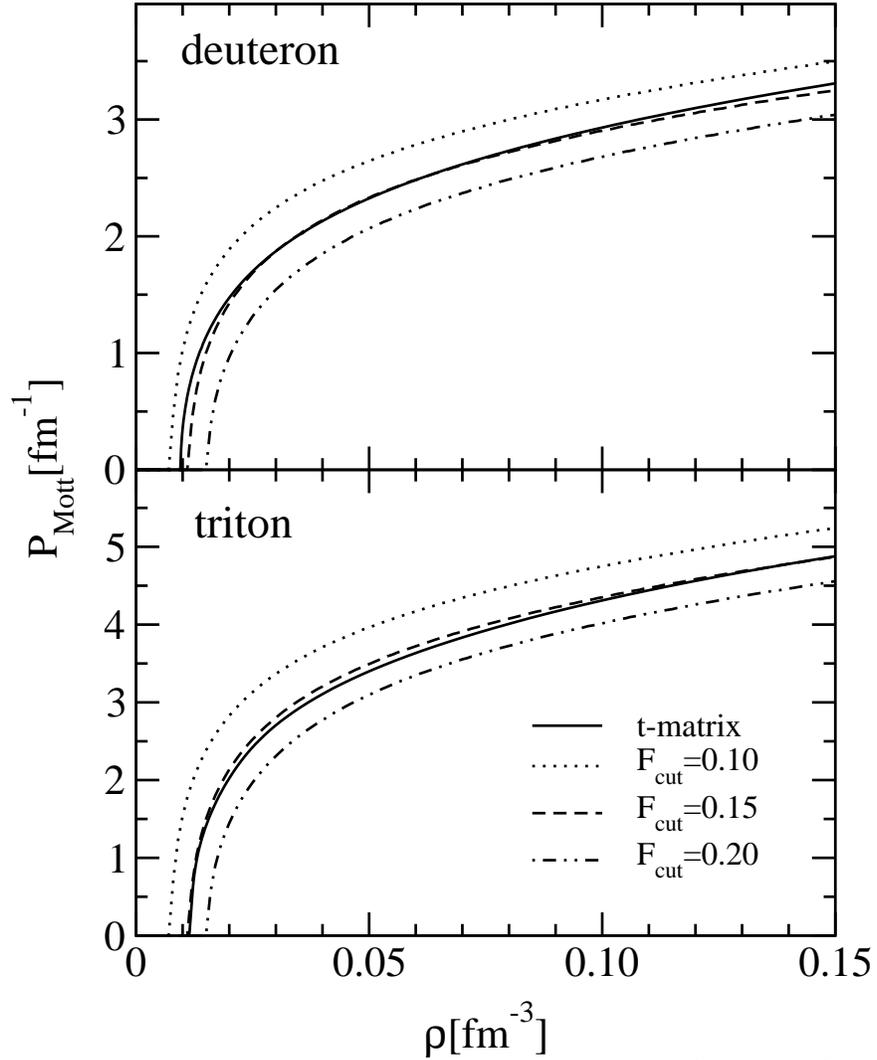}
\caption{\protect\label{fig:deumott} Deuteron and triton Mott momenta
$P_{\rm Mott}$ shown as a function of density $\rho$ at fixed
temperature of $T=10$ MeV. The solid line represents results of
the $t$ matrix approach.  The dashed, dotted and dash-dotted
lines
represent the deuteron Mott momenta from the parameterization given in
Eq.~(\ref{eqn:MOTTMOD}) for three different cut-off values $F_{\rm
cut}$.}  \end{center}
\end{figure}

The Mott effect occurs when a bound state wave function requires too
many momenta components in the regime already occupied by momenta of
other particles (Pauli blocked).  In Ref.~\cite{dan91}
a~geometrical
model was introduced in order to account for the Mott effect in
a~numerical solution of the transport equations~(\ref{eqn:Boltz}).
Specifically, in the model the~formation of a~quasiparticle
cluster is allowed only
if the overlap of the respective isolated bound state wave function
$\phi$ and the Fermi sphere given in Eq.~(\ref{eqn:fermi}) is lower
then a specific cut-off value $F_{\mathrm{cut}}$.  For the deuteron,
the~formation is then allowed if~\cite{dan91}
\begin{eqnarray}\label{eqn:MOTTMOD}
\int d^3\,q\; f(q+\frac{P_{\rm c.m.}}{2}) \; |\phi(q)|^2 \le
F_{\rm cut} \, ,
\end{eqnarray}
where $P_{\rm c.m.}$ denotes the net momentum of the bound state in
the medium frame and $q$ denotes the relative momentum of the nucleons
in the deuteron.  The~cut-off $F_{\rm cut}$ needs to be specified and
it enters as a parameter the collision simulation.  For the clusters
in the Boltzmann equation set, we use the calculated deuteron and
triton Mott momenta~\cite{schadow,ARN} to restrict the choices for the
parameter $F_{\rm cut}$.  Note, however, that the procedure is not
completely stringent since the methods of calculating the Mott
momenta, following the geometrical picture and the $t$ matrix
approach, are different.  The~differently calculated Mott momenta are
shown in in Fig.~\ref{fig:deumott} for the temperature~$T=10$~MeV.  No
bound states are possible for the region below a~respective curves in
Fig.~\ref{fig:deumott}.  As seen in Fig.~\ref{fig:deumott},
the~parameterization given in Eq.~(\ref{eqn:MOTTMOD}) with a cut-off
value of $F_{\rm cut}=0.15$ fits reasonably the microscopic
calculation of the Mott momenta for a~temperature of $T=10$ MeV and
a~wide density range.  Note, that the cut-off values for both
triton/helion and deuteron can be chosen to be approximately the same.
We find only a slight variation of $F_{\rm cut}$ on the temperature
($F_{\rm cut}\simeq0.17$ for $T=5$ MeV). To illustrate sensitivity to
different $F_{\rm cut}$, we have plotted the Mott lines for $F_{\rm
cut}=0.10$ and $F_{\rm cut}=0.20$ in Fig.~\ref{fig:deumott}.  A~{\em
larger} value of $F_{\rm cut}$ extends the space for {\em more} bound
states.  We shall study the influence of the Mott momentum on the
results of reaction simulations.  The~direct implementation of the $t$
matrix calculation for the Mott momenta into the BUU simulation
requires the calculation of a local temperature that is very time
consuming.  Therefore we will make use of the geometrical model, with
a~properly adjusted global value of $F_{\rm cut}$, to account for the
Mott effect in the numerical solution of Eq.~(\ref{eqn:Boltz}).

\begin{figure}[t]
  \begin{center}
\psfig{figure=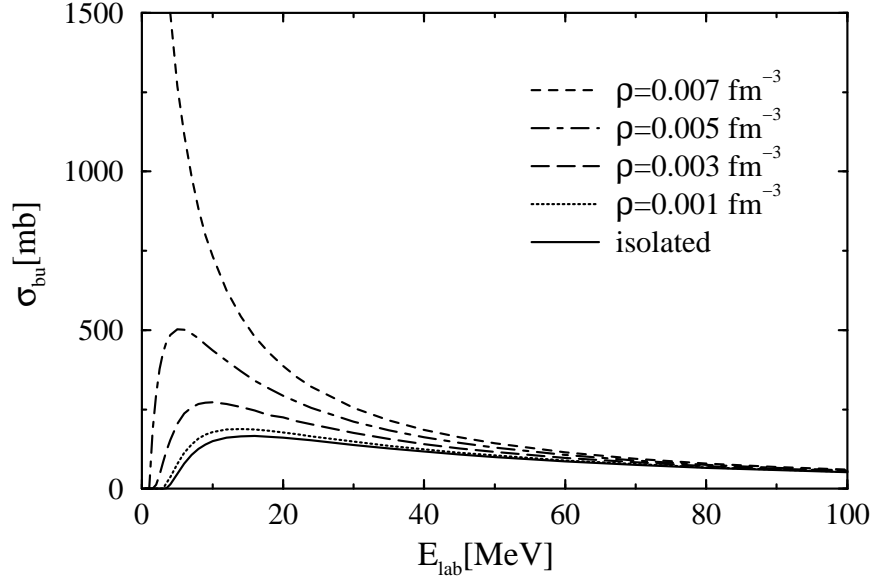,width=0.7\textwidth}
\caption{\protect\label{fig:cross} Neutron-deuteron break-up cross
  section for a three-body system at rest in the nuclear medium. Solid
  line represents isolated break-up cross section. Other lines
are for different   densities at the temperature $T=10$ MeV.}
  \end{center}
\end{figure}

In Eq.~(\ref{eqn:sig0}) we give the relation between the isolated
three-particle transition operators and the total deuteron break up
cross section.  The~same equation is appropriate in defining the
medium-dependent cross-section now in terms of the medium-dependent
transition operator $U$ given in Eq.~(\ref{eqn:AGS2}).  The~total
deuteron break-up cross section can be used in the collision
integrals~(\ref{eqn:react}) and~(\ref{eqn:react2}), as e.g.\ has been
demonstrated in Ref.~\cite{dan91}. For illustration, we show in
Fig.~\ref{fig:cross} the in-medium break-up cross section at a
temperature of $T=10$ MeV for the three-body system at rest in the
nuclear medium, explored to a larger extend
in~\cite{bey96,bey97,kuhrts}. The solid line represents the isolated
cross section that reproduces the experimental break-up data; other
lines correspond to different nuclear matter densities.  The~influence
of the in-medium cross sections on characteristic kinetic quantities
such as the chemical relaxation time is significant as has been shown
in~\cite{kuhrts}.

For a specific heavy ion reaction, we now investigate (i) to what
extend the medium dependence of the deuteron break-up cross section
affects the observables and, in particular, the~deuteron production
and (ii) how sensitive are the observables to the parameterization of
the Mott momenta.  To this end, we consider the central collision
$^{129}$Xe+$^{119}$Sn at E/A=50 MeV and compare our theoretical
results with the experimental data of the INDRA
collaboration~\cite{INDRA}.

\begin{figure}[t]
  \begin{center}
\psfig{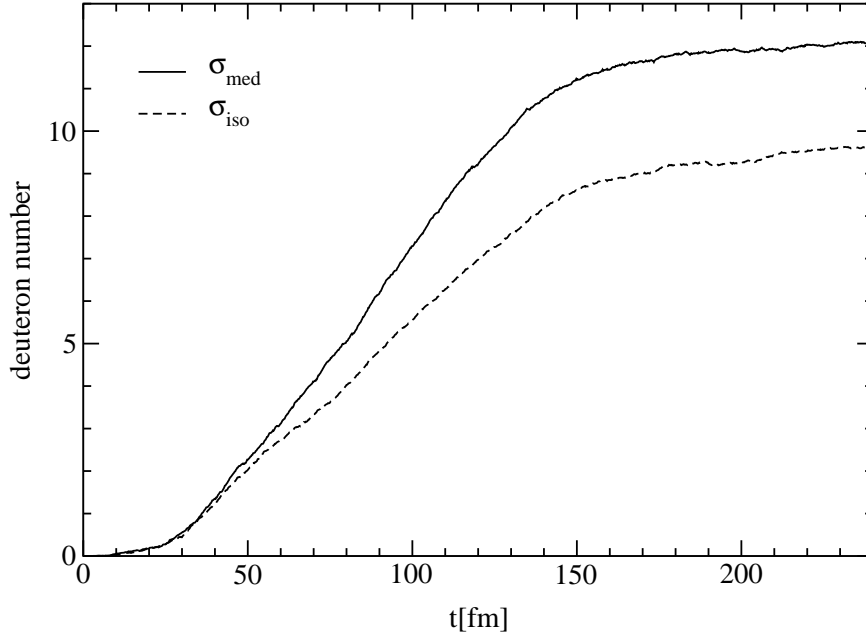}
\caption{\protect\label{fig:number} Integrated deuteron number as a
function of elapsed collision time when utilizing in-medium
(upper curve) and
isolated (lower curve) rates from BUU simulation with $F_{\rm
cut}=0.15$.}
\end{center}
\end{figure}

 To provide a first impression on how the
use of in-medium rates in the Boltzmann collision integrals influences
the outcome of the reaction simulation, Fig.~\ref{fig:number}
shows the total deuteron number
(gain minus loss) as a function of elapsed time.  The~upper
line is for
the in-medium cross sections depending on the local density
and the temperature.  The lower line is for the isolated cross
sections that reproduce the experimental scattering data.
Clearly, the
use of in-medium rates leads to a significant increase in the
total number of deuterons.
In both simulations, we have used $F_{\rm cut}=0.15$. Though
the increase in the deuteron number
may be significant, the theoretical value of the total deuteron
number
may be still too uncertain for a direct comparison to data.
This is because
the effect of the heavier
$A\ge 4$ clusters is not yet included in our simulation.

\begin{figure}[t]
  \begin{center}
\psfig{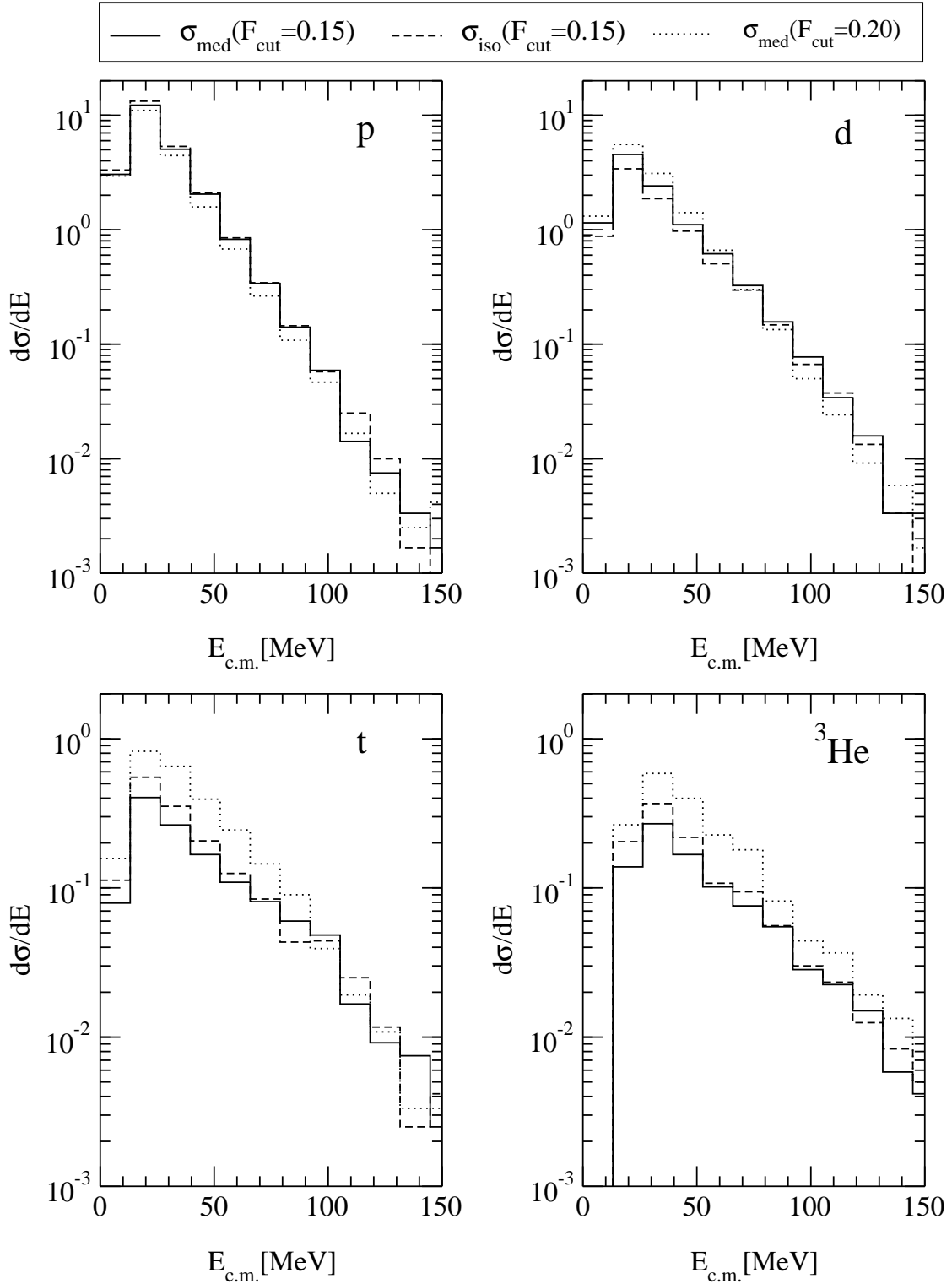}
\caption{\protect\label{fig:exU} Light charged particle spectra in the
  center of mass system for the reaction $^{129}$Xe+$^{119}$Sn at 50
  MeV/A.
Calculations with in-medium $Nd$ reaction rates are represented
by solid lines, while those with isolated $Nd$ break-up cross
sections are represented by dashed lines.
In both cases $F_{\rm cut}=0.15$ is employed.
Calculations
  with $F_{\rm cut}=0.20$ and in-medium reaction rates are represented
  by the dotted lines.}
  \end{center}
\end{figure}

The influence of the different medium effects on the spectral
distribution of proton, deuteron, triton, and helion clusters is shown
in Fig.~\ref{fig:exU}.  Solid line shows the results of our
calculation using the {\em medium-dependent} cross-sections in the
collision integrals of the BUU simulation.  Dashed line shows the
spectra obtained using the {\em isolated} cross sections that
reproduce the experimental data.  In all cases we have included
the Mott effect and for the solid and the dashed lines we use
$F_{\rm
cut}=0.15$.  To demonstrate the sensitivity to different Mott
momenta, the dotted line shows results obtained using $F_{\rm
cut}=0.20$ for medium-dependent rates. As explained in
Ref.~\cite{bey96,bey97,kuhrts}, the~deuteron break-up cross section is
strongly enhanced near the threshold.  As a~consequence, we find
an~enhancement of about 30\% in the deuteron number in the energy
range $E_{\rm c.m.}\le 50$ MeV if we compare the dashed line (for
isolated deuteron break-up cross-sections) to the solid line (for
medium-dependent deuteron break-up cross sections).  A~larger
absorption implies a~larger production rate and maintaining the
possible chemical equilibrium down to lower freeze-out densities.
Inspecting the dotted line, we find, as~expected, that the~number of
clusters increases with the rise in~$F_{\rm cut}$.  Also,
simultaneously, the~spectra appear steeper.

\begin{figure}[t]
  \begin{center}
\psfig{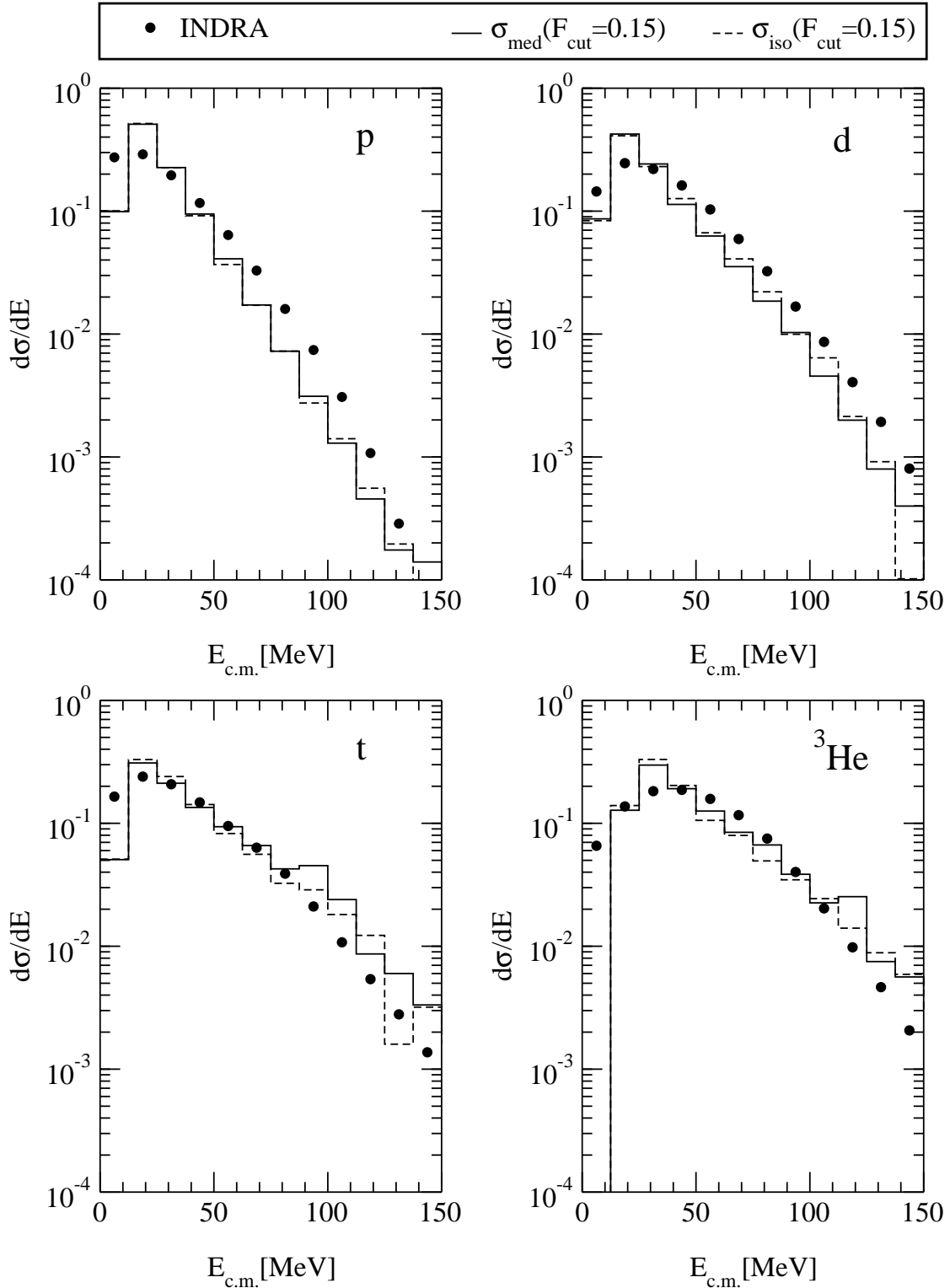}
\caption{\protect\label{fig:exR} Renormalized light charged light
particle spectra in the center of mass system for the reaction
$^{129}$Xe+$^{119}$Sn at 50~MeV/A.  The filled circles represent the
data of the INDRA collaboration~\protect\cite{INDRA}.  The solid line
shows the calculations with the in-medium $Nd$ reaction rates, while
the~dashed line shows a calculation using the isolated $Nd$ break-up
cross section; both with $F_{\rm cut}=0.15$.}  \end{center}
\end{figure}

Figure~\ref{fig:exR} compares energy spectra of the light $A\le 3$
clusters to INDRA data renormalized as in Ref.~\cite{INDRA}.  To
compare to our calculations we renormalize our results in the same
fashion as data, i.e. the areas below the respective curves are
normalized to  the same fixed value.  As before the results
obtained with the {\em
medium-dependent} cross-sections are represented by a solid line;
the~dashed line shows the results of the coupled BUU calculation using
the {\em isolated} deuteron break-up cross sections.  In~both
calculations we use $F_{\rm cut}=0.15$.
Overall, the renormalization seems
to reduce the effect of using different elementary rates in the
collision integrals.  For the deuteron, the {\em in-medium} rates lead
to a marginally steeper shape.  Differences at higher c.m.\ energies
can be attributed to statistical fluctuations.

Considering now the three-nucleon clusters, we find the helion/triton
energy spectrum in a reasonable agreement with the experimental data.
The~only in-medium effect that we consider for the three-nucleon
clusters is the Mott effect.  A~possible modification of the
three-particle cluster formation and break-up is disregarded.
The~observed reduction, see Fig.~\ref{fig:exU}, of the three-nucleon
cluster production in the calculation using the medium deuteron
break-up cross-section is a~response to the enhanced deuteron
production.

\begin{figure}[t]
  \begin{center}
\psfig{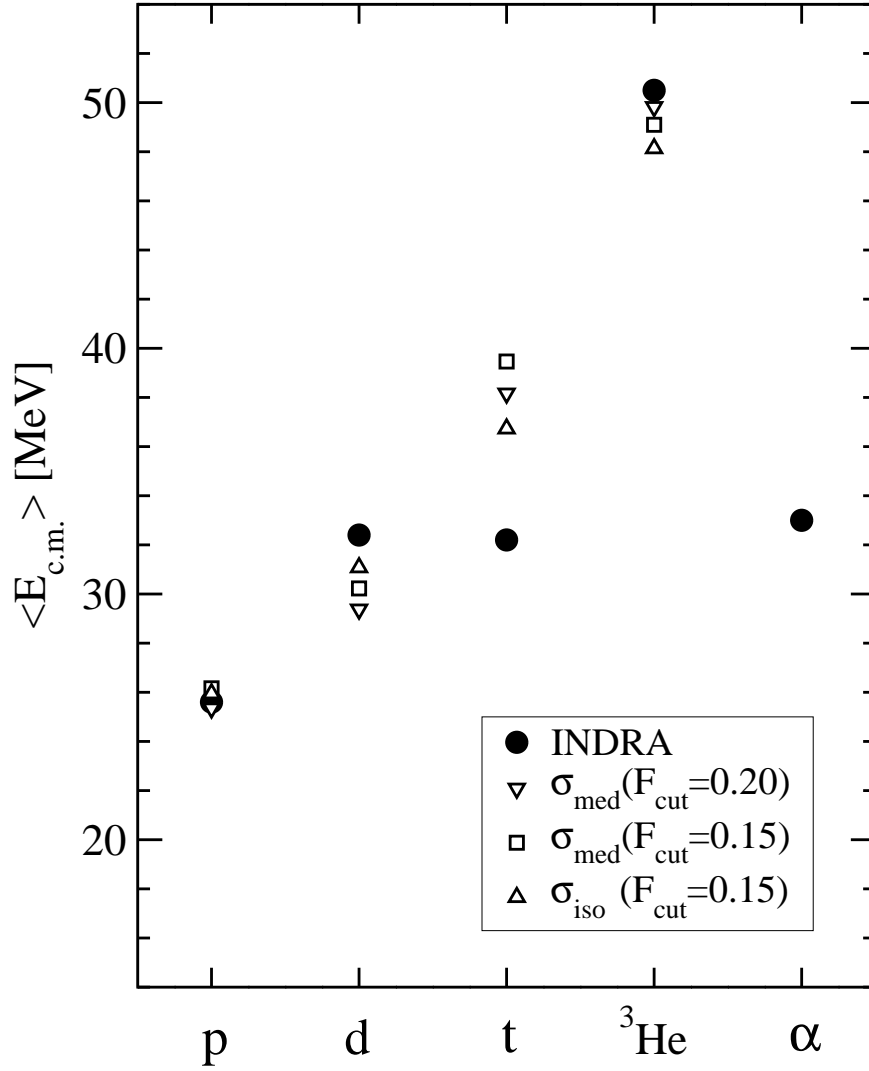}
\caption{\protect\label{fig:Emean} Mean transverse energy of
  light charged fragments in the angular range of $-0.5\le
  \cos\theta_{c.m.}\le 0.5$. }
  \end{center}
\end{figure}

Finally, in Fig.~\ref{fig:Emean} we show the measured values (filled
circles) for the mean energies of the light clusters emitted in the
transverse direction. While the~overall tendency is well
reproduced,
the~triton energy shows some discrepancy.  The~medium effects are less
pronounced in this observable.  The~triton energy is only slightly
reduced with an enhancement in the parameter~$F_{\rm cut}$.  No
theoretical results are given for
$\alpha$-particles which are not yet included in the
simulations.

\section{Conclusion}
Within the microscopic transport description of heavy-ion
collision dynamics, we have demonstrated the influence of
medium effects
on some typical observables.  The~dominant medium effects are
self-energy and statistical corrections, i.e.\ for clusters the Mott
effect, and the change of reaction rates that lead to faster time
scales.  The magnitude of the in-medium effects depends on the density and
the energy deposited in the system.  The basis for microscopic
calculations of the in-medium effects is the cluster mean-field
expansion or Dyson equation approach.  The~effective few-body
equations resulting from these approaches are numerically solved using
well established and controllable few-body techniques.  The~chosen
example of a~heavy-ion reaction is $^{129}$Xe+$^{119}$Sn at beam
energy of 50 MeV/A.  Both effects, Mott and rate modification,
affect
the considered observables in a~comparable fashion.  Presently, we
have implemented the modifications of rates for the three-body
break-up and formation only. However, through the coupling of the
Boltzmann equations, the~number of three-nucleon clusters is also
affected.  We argue that a~complete treatment requires a~similar
approach for the three- and four-nucleon clusters before a~decisive
comparison with experiments can be made. First calculations for the
Mott effect of the $\alpha$-particle solving a proper four-body
equation has been given in~\cite{alpha}.

\section{Acknowledgment}
 This work has been supported by the Deutsche Forschungsgemeinschaft
and by the National Science Foundation under Grant PHY-0070818. MB and
CK thank the NSCL at MSU for the warm hospitality extended to them
during their respective stays.

\end{document}